# Hiding Image in Image by Five Modulus Method for Image Steganography

Firas A. Jassim

**Abstract**— This paper is to create a practical steganographic implementation to hide color image (stego) inside another color image (cover). The proposed technique uses Five Modulus Method to convert the whole pixels within both the cover and the stego images into multiples of five. Since each pixels inside the stego image is divisible by five then the whole stego image could be divided by five to get new range of pixels 0..51. Basically, the reminder of each number that is not divisible by five is either 1,2,3 or 4 when divided by 5. Subsequently, then a 4×4 window size has been implemented to accommodate the proposed technique. For each 4×4 window inside the cover image, a number from 1 to 4 could be embedded secretly from the stego image. The previous discussion must be applied separately for each of the R, G, and B arrays. Moreover, a stego-key could be combined with the proposed algorithm to make it difficult for any adversary to extract the secret image from the cover image. Based on the PSNR value, the extracted stego image has high PSNR value. Hence this new steganography algorithm is very efficient to hide color images.

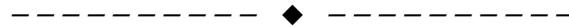

## 1 INTRODUCTION

In past decade, the concept of steganography had obtained a wide attention in the filed of information security. The word steganography comes from the Greek words stegos which means covered and graphia which means writing. Steganography refers to the process of hiding secret message inside any kind of media such as audio, video, or photo [1]. The basic idea in almost data hiding systems is the weakness of human perceptual such as vision, listening, and hearing. There is a common confusion between steganography and cryptography because they both are used to protect secret information. The main difference between Steganography and cryptography is that, cryptography concentrates on keeping the contents of a message secret while steganography concentrates on keeping the existence of a message secret [2]. Therefore, to add more protection to the secret information both steganography and cryptography are needed but each with its own technique. Subsequently, if the idea is to hide the appearance of the secret message, then the procedure of steganography will arise [2]. Combining both steganography and cryptography in the same secret message, these will double the job for the third party to extract the stego image from the cover image and another job is to decipher the encrypted message [3].

Moreover, steganography is not the same as watermarking [4]. The main dissimilarity between steganography and watermarking is the absence of an opponent. In watermarking applications like copyright protection and authentication, there is an active opponent that would attempt to remove, abolish or forge watermarks. In steganography there is no such active opponent as there is no value associated with the act of removing the information hidden in the content [5]. Several methods have been proposed for image based steganography, the simplest and widest known steganographic method is the Least Significant Bit (LSB) which replaces the least significant bits of pixels selected to hide the information. A detailed discussion about LSB could be found in [6]. Also, there are a wide variety of different techniques with their own advantages and disadvantages were constructed in steganography. Research in hiding data inside image using steganography technique has been done by many researchers [7][8][9]. Also, an excellent theoretical background about steganography could be found in [10].

The paper is organized as follows: In the next section, an overview of Five Modulus Method was presented. The proposed steganography algorithm was presented in Section 3. The analysis of stego-key was discussed in section 4. Moreover, experimental results and conclusions are presented in Sections 5 and 6, respectively.

## 2 FIVE MODULUS METHOD

Five Modulus Method (FMM) was firstly originated by [11]. Basically, it was founded as a method for image compression. The basic idea behind FMM is to transform the whole image into multiples of five. Since the human eye does not differentiate between the original image and the transformed FMM image [11]. It is known that, for each of the R, G, and B arrays in the color image are consist of pixel values varying from 0 to 255. The main impression of FMM is to transform the whole pixels inside the image into numbers divisible by five. Steganography algorithms concentrate on introducing as small deformation in the cover image as possible [12]. According to figure (1), it is clear that the FMM transformation does not affect the Human Visual System (HVS).

---

• *Firas A. Jassim is with the Management Information Systems Department, Irbid National University, Irbid-Jordan.*

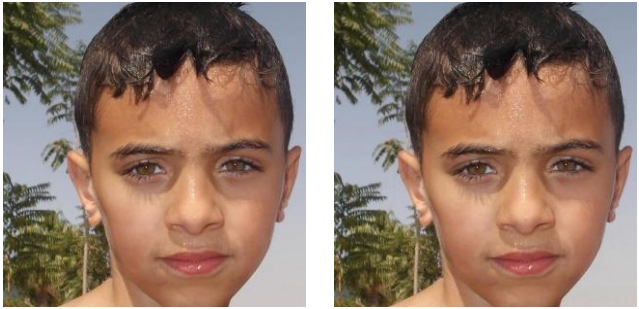
Fig. 1. Original image (Saif.bmp) (left) FMM image (right)

| 17 | 44 | 12 | 2 |
|----|----|----|----|
| 37 | 87 | 101 | 207 |
| 189 | 220 | 121 | 23 |
| 93 | 99 | 170 | 8 |

…

| 15 | 45 | 10 | 0 |
|----|----|----|----|
| 35 | 85 | 100 | 205 |
| 190 | 220 | 120 | 25 |
| 95 | 100 | 170 | 10 |

Fig. 2. Original block (left) FMM block (right)

According to figure (2), the new pixel value in the k×k window will be a multiple of five as: 0, 5, 10, 20, …, 100, 105, …, 200, 205, 210, …, 250, 255. Hence, the values that are not divisible by 5 are distinct inside k×k block. Now, the good thing is that by dividing the new values by 5. A new range of values will be constructed as: 0, 1, 2, 3, …, 51. Since 0÷5=0, 5÷5=1, 10÷5=2, …, 250÷5=50, 255÷5=51. Hence, the new range is consisting of 52 distinct values.

## 3 PROPOSED ALGORITHM

The proposed technique is based on the FMM transformation. Therefore, all the pixels inside the FMM images are all multiples of 5 only. Basically, the original pixel values are in the range 0 to 255 for each of the Red, Green, and Blue arrays. After the FMM transform, the new range is still 0 to 255 but with multiples of 5 only, i.e. 0, 5, 10, 15, …, 255. The FMM method will be applied for both the cover and the stego images. Now, if we divide the new range by 5 we can get a simpler range that is 0, 1, 2, …51, i.e. 0÷5=0, 255÷5=51. The model of the proposed stegosystem could be introduced as a scheme in figure (3).

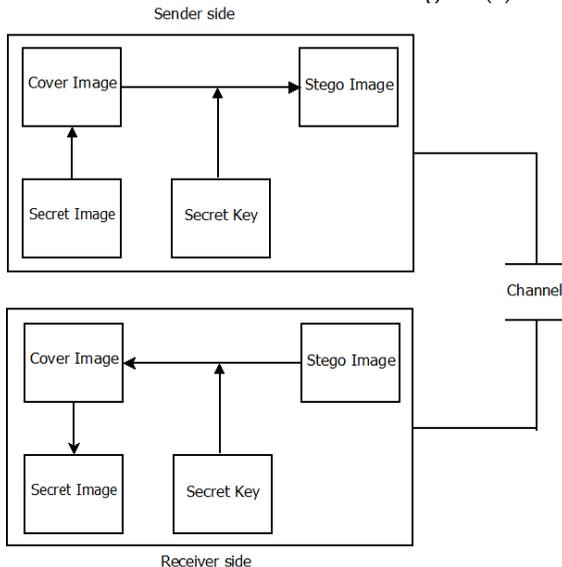
Fig 3. The model of the proposed stegosystem

### 3.1 Window Size Determination

Since the reminder of each number that is not divisible by five is either 0, 1,2,3 or 4. The zero reminder means that it is divisible by 5 while 1, 2, 3, 4 reminders means the opposite. The determination of the suitable window size is based on two concepts. The first one is that the new range values are 0..51, i.e. 52 distinct values. The second one is that the reminders of numbers that are not divisible by five are 1, 2, 3, and 4. Since the selection of window is based on the square criterion which means k×k. Then, nearest number which can accommodate the 52 values is 64. But the square root of 64 is 8 which means 8×8 window size and that is large. Now, to reduce the large window size, the second concept which is the 1, 2, 3, and 4 reminders may be applied. Since the whole values are multiples of five, then each number that is not divisible by five will be inconsistent with the others inside window. Therefore, by dividing 64 by 4 we get 16. Subsequently, a 4×4 window size that contains 16 items will be suitable to accommodate the proposed algorithm. A 4×4 window could holds the 52 values by using a simple looping method.

### 3.2 Looping Method

As stated in the previous section, the nearest upper integer to encompass the 52 values is 64 and also the suitable window size is 4×4 which contains 16 position. The new range of values 0..51 could be apportioned into four parts each one consist of 16 elements. The first part contains the first 16 values which are 0..15 (could be named as P1), while the second part varies 16..31 (P2), and the third one is 32..47 (P3). Finally, since the last range is 48..51 which consists of only four numbers but there is no doubt it need a stand alone fourth part (P4). The looping method consists of adding the suitable reminder to the pixel inside the window. If the position of the stego pixel is less than 15, i.e. in the first part (P1) then 1 may be added to its position inside block, 2 to P2, 3 to P3, and 4 to P4.

TABLE 1
Arbitrary 4×4 windows

| 26 | 25 | 25 | 25 | 25 | 25 | 25 | 25 | 26 | 25 | 25 | 25 |
|----|----|----|----|----|----|----|----|----|----|----|----|
| 25 | 25 | 25 | 25 | 27 | 25 | 25 | 25 | 25 | 25 | 25 | 25 |
| 25 | 25 | 25 | 25 | 25 | 25 | 25 | 25 | 25 | 25 | 28 | 25 |
| 25 | 25 | 25 | 25 | 25 | 25 | 25 | 25 | 25 | 25 | 25 | 25 |

Concerning the first 4×4 window, the only number that is not divisible by 5 is 26. The position of 26 is in the first index of the 4×4 window. The first index holds the value of 0 since P1 indexes are 0..15. Hence, stego pixel inside 26 is 0×5=5. The multiplication of 5 was done because the aim is to retrieve the original value which was divided by 5 previously. Moreover, in the second 4×4 window, the reminder of 27 is 2 when divided by 5 and this means two loops, i.e. this pixel lies in P2 where its range varies 16..31. The position of 27 in the second window using vertical tracing is 2. Therefore, the second stego pixel inside P2 is 17×5=85. The number 17 is the second number in the range 16..31. Finally, for the third window, the reminder of 28 is 3 which mean that it is in P3, i.e. three loops. Since the range for P3 is 32...47, and its position is 11 vertically, then the stego pixel inside the third window is 41×5=205. The value of 41 was used as the eleventh index in the range 32..47.

## 4 PRIVATE STEGO-KEY

A stego-key is used to control the hidden information in such away that helps to recover the secret message. Stego-key encodes information in such a way that the third party cannot extract the stego image unless he holds the stego-key. A stego-key is a password, which ensures that only the recipient who knows the stego-key has the ability to extract the stego image from the cover image. Furthermore, combining the secret key with the proposed algorithm makes it difficult to extract the stego image by any adversary.

According to figure (4), the technique used to produce a stego-key is by forward shifting either horizontally or vertically. As mentioned previously, a 4×4 window contains 16 positions, therefore; by shifting the position of the stego pixel by n positions inside the same 4×4 window. The shifting procedure is done either horizontally or vertically depending on the first digit in the stego-key. Furthermore, if an adversary looks forward to extract the stego image then he would try tremendous number of contingencies. For each 4×4 window there are 16 positions for both the horizontal and the vertical tracing. Hence, in a 512×512 image, the exact number of trials are $(16 \times 2)^{128}$ which equals to $4.5 \times 10^{192}$ where 128 comes from 512÷4 which are the total number of 4×4 windows in the 512×512 cover image and 2 for both the horizontal and the vertical tracing.

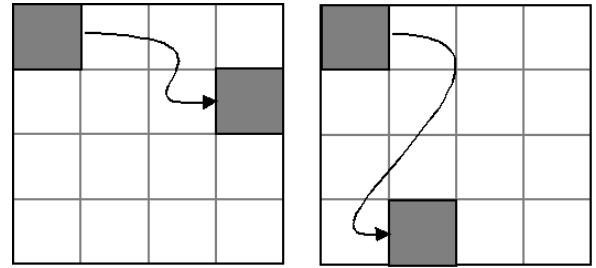

Fig. 4. Horizontal shift by seven bits (left) Vertical shift by seven bits (right)

In order to determine the direction of shifting, the starting digit of the stego-key will be saved for this purpose. Hence, a value of zero in the first digit means that the direction is horizontal while the value of one means vertical. To illustrate this point of view, suppose the password 791432 is to be used as a stego-key for an arbitrary block of 4×4 window pixels, table (2). By adding the value of 0 to the left of the stego-key, i.e. 0791432, then a horizontal direction will be used, table (3). On the other hand, if the value of the first digit is 1, i.e. 1791432, this means vertical direction, table (4).

TABLE 2
Arbitrary block of 4×4 windows

| 21 | 20 | 20 | 20 | 21 | 20 | 20 | 20 | 21 | 20 | 20 | 20 | 21 | 20 | 20 | 20 | 20 | 20 | 20 | 20 | 20 | 20 | 20 | 20 |
|----|----|----|----|----|----|----|----|----|----|----|----|----|----|----|----|----|----|----|----|----|----|----|----|
| 20 | 20 | 20 | 20 | 20 | 20 | 20 | 20 | 20 | 20 | 20 | 20 | 20 | 20 | 20 | 20 | 21 | 20 | 20 | 20 | 21 | 20 | 20 | 20 |
| 20 | 20 | 20 | 20 | 20 | 20 | 20 | 20 | 20 | 20 | 20 | 20 | 20 | 20 | 20 | 20 | 20 | 20 | 20 | 20 | 20 | 20 | 20 | 20 |
| 20 | 20 | 20 | 20 | 20 | 20 | 20 | 20 | 20 | 20 | 20 | 20 | 20 | 20 | 20 | 20 | 20 | 20 | 20 | 20 | 20 | 20 | 20 | 20 |

TABLE 3
Shifting the stego-pixels horizontally by password (0791432)

| 20 | 20 | 20 | 20 | 20 | 20 | 20 | 20 | 20 | 21 | 20 | 20 | 20 | 20 | 20 | 20 | 20 | 20 | 20 | 20 | 20 | 20 | 20 | 20 |
|----|----|----|----|----|----|----|----|----|----|----|----|----|----|----|----|----|----|----|----|----|----|----|----|
| 20 | 20 | 20 | 21 | 20 | 20 | 20 | 20 | 20 | 20 | 20 | 20 | 21 | 20 | 20 | 20 | 20 | 20 | 20 | 20 | 21 | 20 | 21 | 20 |
| 20 | 20 | 20 | 20 | 20 | 21 | 20 | 20 | 20 | 20 | 20 | 20 | 20 | 20 | 20 | 20 | 20 | 20 | 20 | 20 | 20 | 20 | 20 | 20 |
| 20 | 20 | 20 | 20 | 20 | 20 | 20 | 20 | 20 | 20 | 20 | 20 | 20 | 20 | 20 | 20 | 20 | 20 | 20 | 20 | 20 | 20 | 20 | 20 |

TABLE 4
Shifting the stego-pixels vertically by password (1791432)

| 20 | 20 | 20 | 20 | 20 | 20 | 20 | 20 | 20 | 20 | 20 | 20 | 20 | 21 | 20 | 20 | 20 | 20 | 20 | 20 | 20 | 20 | 20 | 20 |
|----|----|----|----|----|----|----|----|----|----|----|----|----|----|----|----|----|----|----|----|----|----|----|----|
| 20 | 20 | 20 | 20 | 20 | 20 | 21 | 20 | 21 | 20 | 20 | 20 | 20 | 20 | 20 | 20 | 20 | 20 | 20 | 20 | 20 | 20 | 20 | 20 |
| 20 | 20 | 20 | 20 | 20 | 20 | 20 | 20 | 20 | 20 | 20 | 20 | 20 | 20 | 20 | 20 | 20 | 20 | 20 | 20 | 20 | 20 | 20 | 20 |
| 20 | 21 | 20 | 20 | 20 | 20 | 20 | 20 | 20 | 20 | 20 | 20 | 20 | 20 | 20 | 20 | 21 | 20 | 20 | 20 | 21 | 20 | 20 | 20 |

According to table (3), the first 4×4 window contains the number 21 as non divisible by 5 and that is a stego pixel. If and adversary hops to implement the proposed algorithm without the stego key, then huge number of possibilities could be tried. Alternatively, as the stego key used in table (3) is 0791432, then the first window should be back shifted with 7 horizontal numbers. Hence, the receiver could easily extract the stego image with the help of the stego key. The same procedure could be implemented to table (4) but with vertical tracing.

## 5 EXPERIMENTAL RESULTS

In order to demonstrate the proposed steganography algorithm, four bitmap test images (Shaikh1.bmp, Shaikh1.bmp, Shaikh1.bmp, and Shapes.bmp) were used as cover images. Alternatively, four stego images (Stego1.bmp, Stego2.bmp, Stego3.bmp, and Stego4.bmp) were used as stego images with the previous four cover images, respectively, figure (5). As stated previously, a 4×4 window size was used to implement the proposed algorithm. Also, for each 4×4 window, one stego pixel could be embedded inside each 4×4 window. Hence, the maximum payload could be reached by the proposed method is always 25% for all image sizes.

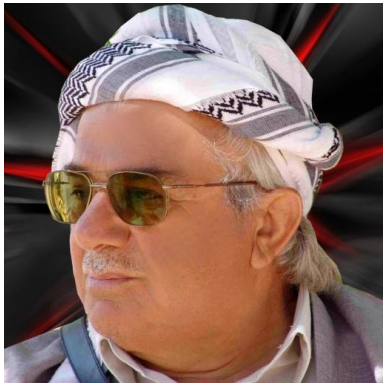

(shaikh1.bmp) 512×512

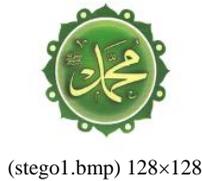

(stego1.bmp) 128×128

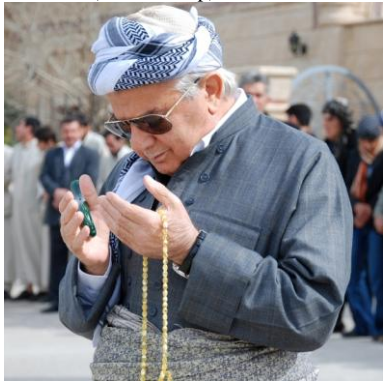

(shaikh2.bmp) 1024×1024

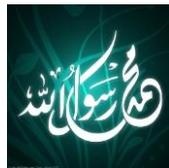

(stego2.bmp) 256×256

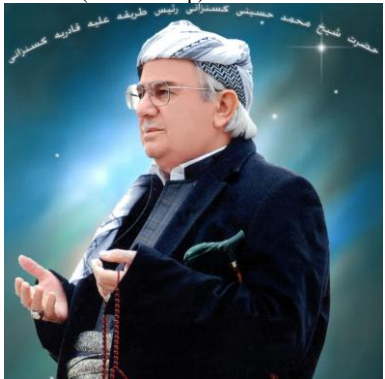

(shaikh2.bmp) 2048×2048

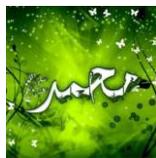

(stego3.bmp) 512×512

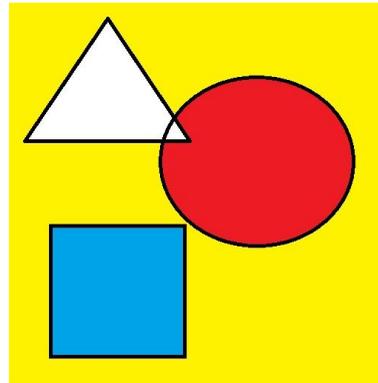

(shapes.bmp) 512×512

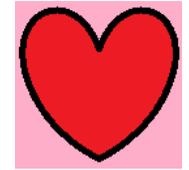

(heart.bmp) 128×128

Fig. 5. Cover images (left) stego images (right)

Moreover, it clearly shown in table (5) that the PSNR values for both the stego and cover images are approximately high (around forties).

TABLE 5
PSNR (dB) values after the stegoanalysis

| Shaikh1 (512×512) | 45.8751 |
|---|---|
| Shaikh2 (1024×1024) | 47.0648 |
| Shaikh3 (2048×2048) | 46.9422 |
| Shapes (512×512) | 41.9158 |
| stego1 (128×128) | 38.3058 |
| stego1 (256×256) | 44.1497 |
| stego1 (512×512) | 47.6531 |
| Stego4 (128×128) | 47.9789 |

## 6 CONCLUSION

In this paper, a novel steganography algorithm has been introduced. The novel algorithm based on the Five Modulus method to reduce the original pixels range from 0..255 into 0..51. The constructed stego images have no noticeable dissimilarities between the original stego images. This method also distributes the stego image uniformly in the cover image. A stego-key may also be combined with the cover image to complicate the way for any adversary who aims to extract the secret image. An error metrics, especially PSNR, have been calculated to prove that the dissimilarities between the original image and the cover images are small and that is based on the high PSNR values.

**Firas A. Jassim** received the BS degree in mathematics and computer applications from Al-Nahrain University, Baghdad, Iraq in 1997, and the MS degree in mathematics and computer applications from Al-Nahrain University, Baghdad, Iraq in 1999 and the PhD degree in computer information systems from the university of banking and financial sciences, Amman, Jordan in 2012. Now, he is working as an assistant professor with the Management Information Systems Department at Irbid National University, Irbid, Jordan. His research interests are Image processing, image compression, image enhancement, image interpolation, image steganography and simulation.